\documentclass[12pt,a4paper]{article}
\usepackage{graphicx}
\pdfoutput=1

\graphicspath{{plots/}}

\def\wg{$W\gamma$ }
\def\baur{\textsc{Baur Wgamma\_Nlo }}
\def\pythia{\textsc{Pythia }} 
\def\ptg{$p_{T}^{\gamma}$ } 

\begin{document}

\thispagestyle{empty}

\baselineskip 18pt
\vspace*{-1in}
\renewcommand{\thefootnote}{\fnsymbol{footnote}}

\begin{flushright}
MCnet/10/02 \\ 
\end{flushright}

\vskip 65pt

\begin{center}
{\Large \bf A study of a NLO matrix element generator for \wg and matching scheme for NLO events and \pythia parton shower} 
\footnote{An abbreviated version to appear in the proceedings of Physics at TeV Colliders, Les Houches 8-26 June 2009} 
\footnote{Work was supported by the European Union Marie Curie Research Training Network MCnet, under contract \makebox{MRTN-CT-2006-035606}}
\\
\setcounter{footnote}{0}
\vspace{10pt}
{\large\bf \large D. Majumder~$^{1}$, K. Mazumdar~$^1$, T. Sj\"ostrand~$^2$} \\ 
\vspace{10pt} 
{\bf $^1$Tata Institute of Fundamental Research, \\ Homi Bhabha Road, Mumbai 400 00
5, India.\\
$^2$Department of Theoretical Physics, Lund University, \\ S\"olvegatan14A, S-223~62, Sweden.} 
\end{center}

\vspace{\fill}

%
%

\begin{abstract}
We present here a study of event generators for the hadronic production of \wg. We compare a NLO \wg matrix element generator with the leading order calculation in \pythia.
A matching scheme between a next-to-leading order \wg matrix element generator by Baur et. al. and the \pythia parton shower is presented. The NLO package produces \wg+0 jet and \wg+1jet final states in the hard scattering and the objective is to consistently match these  to the initial state radiation from \pythia parton shower. The proposed methodology preserves both the rate of the hard scattering process as well as various kinematic distributions of experimental interest.
\end{abstract}

\vskip12pt
\noindent

\vfill
\clearpage
\setcounter{page}{2} 

\section{Introduction}

Even though the Electroweak Theory of the Standard Model (SM) has been tested to remarkable precision in previous and current accelerator-based experiments, the non-Abelian nature of the Electroweak gauge bosons remain to be tested to greater precision. This requires a study of the couplings among the vector bosons, the {\it Triple Gauge Couplings} and the {\it Quartic Gauge Couplings} at higher energy scales and the LHC provides an opportunity to study the bosonic self-interactions. The cross-section of some of the diboson productions, e.g. that of \wg is large enough for a study of $pp\rightarrow W\gamma$ to be feasible at the low energy early running of the LHC with data corresponding to $\sim$100~pb$^{-1}$. In preparation for analysis of data from the LHC, we present here a study of event generators for simulating $pp\rightarrow W\gamma$ event production at the LHC. 

There are broadly two classes of event generators: one which uses exact matrix element calculation up to a fixed order and are used for processes involving a few high particles (matrix element or ME generators) and the other which uses an approximate scheme to generate events up to all orders and involving many particles, most of which are relatively soft and collinear (parton shower or PS generators). Event generation also involves a hadronization procedure to combine partons from hard scattering and underlying events into hadrons which are observable in the experiments. For simulating the hard interaction which produces a W and a $\gamma$ in proton-proton collision, a matrix element calculation is preferred over a parton shower generator. Nevertheless, the parton shower is required for providing the countless softer emissions that accompany a hard scattering as well as simulating the underlying events and finally hadronizing the product quarks and gluons into colour-singlet hadrons, thereby providing a complete picture of an actual event. 

The most popular event generators until now have been of leading order accuracy. However, at the LHC, we believe that higher order corrections will become important and may lead to event topologies very different from those seen at previous hadron colliders. In view of this, we explore the possibility of using a next-to-leading order (NLO) event generator for the \wg production process. We choose \baur \cite{Baur:1993ir} as a dedicated NLO matrix element generator and compare it to \pythia \cite{Sjostrand:2006za} which is a common general purpose leading order generator. \pythia also provides a showering and hadronization mechanism for producing the full event. 

The photon transverse momentum spectrum (\ptg) is the most important observable parameter in the study of \wg production as this variable is most sensitive to the nature of the WW$\gamma$ vertex coupling. We make a detailed study of \ptg from the various source: $q\bar{q'}\rightarrow W\gamma$ hard scattering; photons from initial state radiation from the incoming quarks (ISR); photons from the final state radiation from charged leptons from W-decay (FSR); and photons from an event with multiple hard scattering, e.g. with $q\bar{q'}\rightarrow W\gamma$ and $q\bar{q}\rightarrow g\gamma$ occurring in the same proton-proton collision. 

Eventually we propose a scheme for combining events from the NLO calculation of \baur with the parton shower (which is at leading order) from \pythia which takes into account the "double counting problem" of matching a NLO calculation with a leading order shower. 

\section{\wg production at hadron colliders} 

\begin{figure}[!htbp] 
\begin{center}
\includegraphics[width=0.3\textwidth,height=0.25\textwidth]{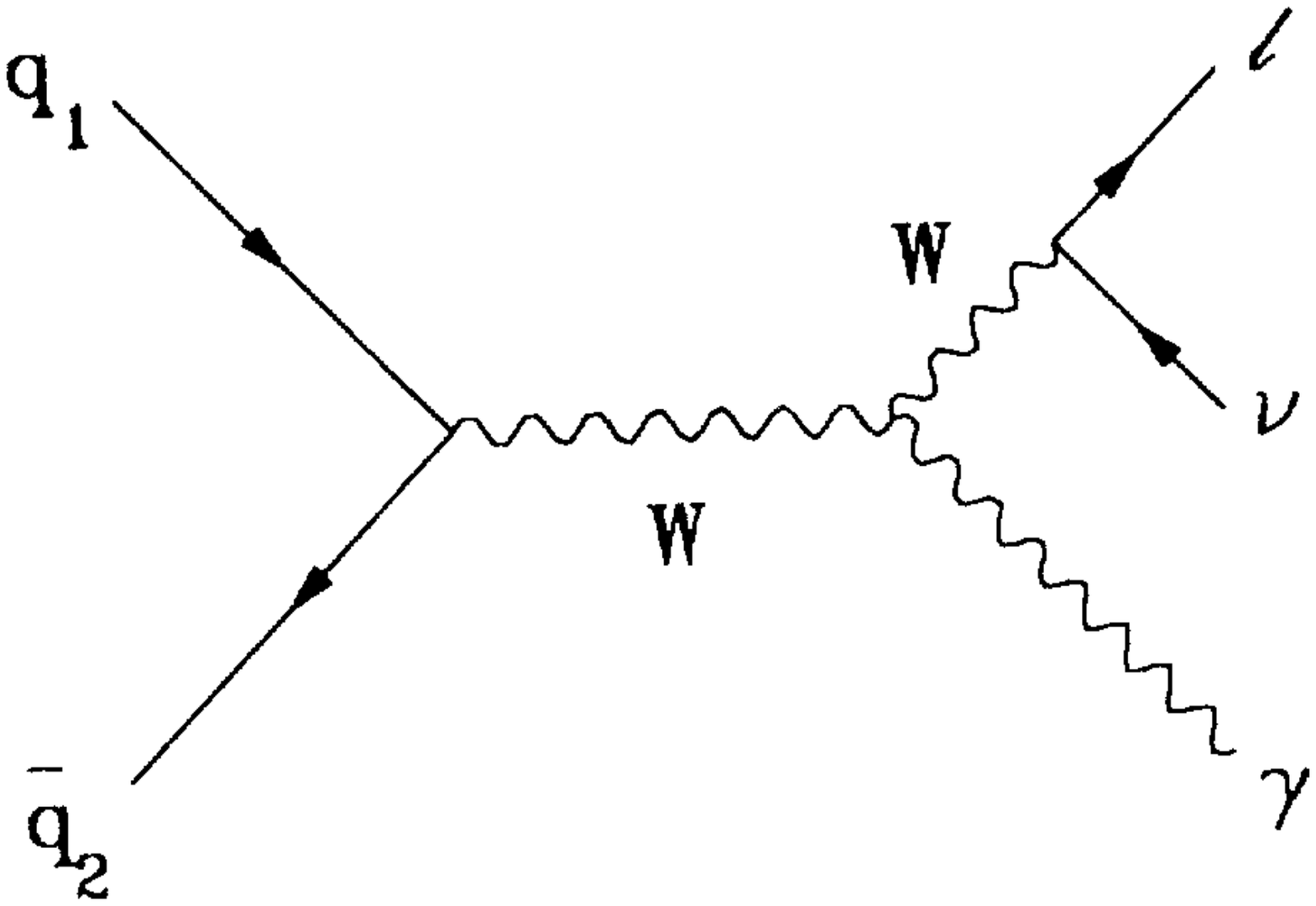}
\hspace{0.1cm}
\includegraphics[width=0.3\textwidth,height=0.25\textwidth]{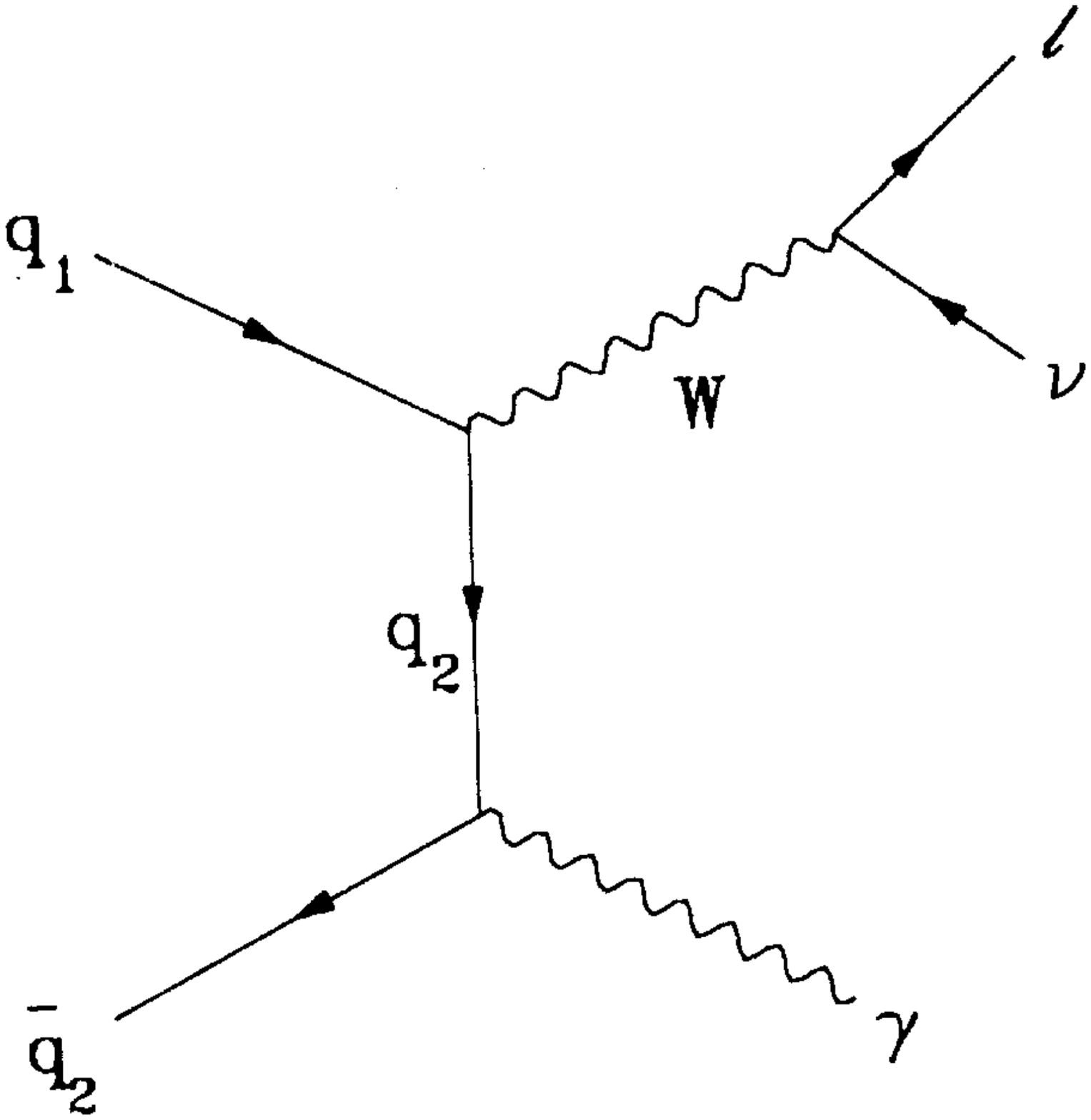}
\hspace{0.1cm}
\includegraphics[width=0.3\textwidth,height=0.25\textwidth]{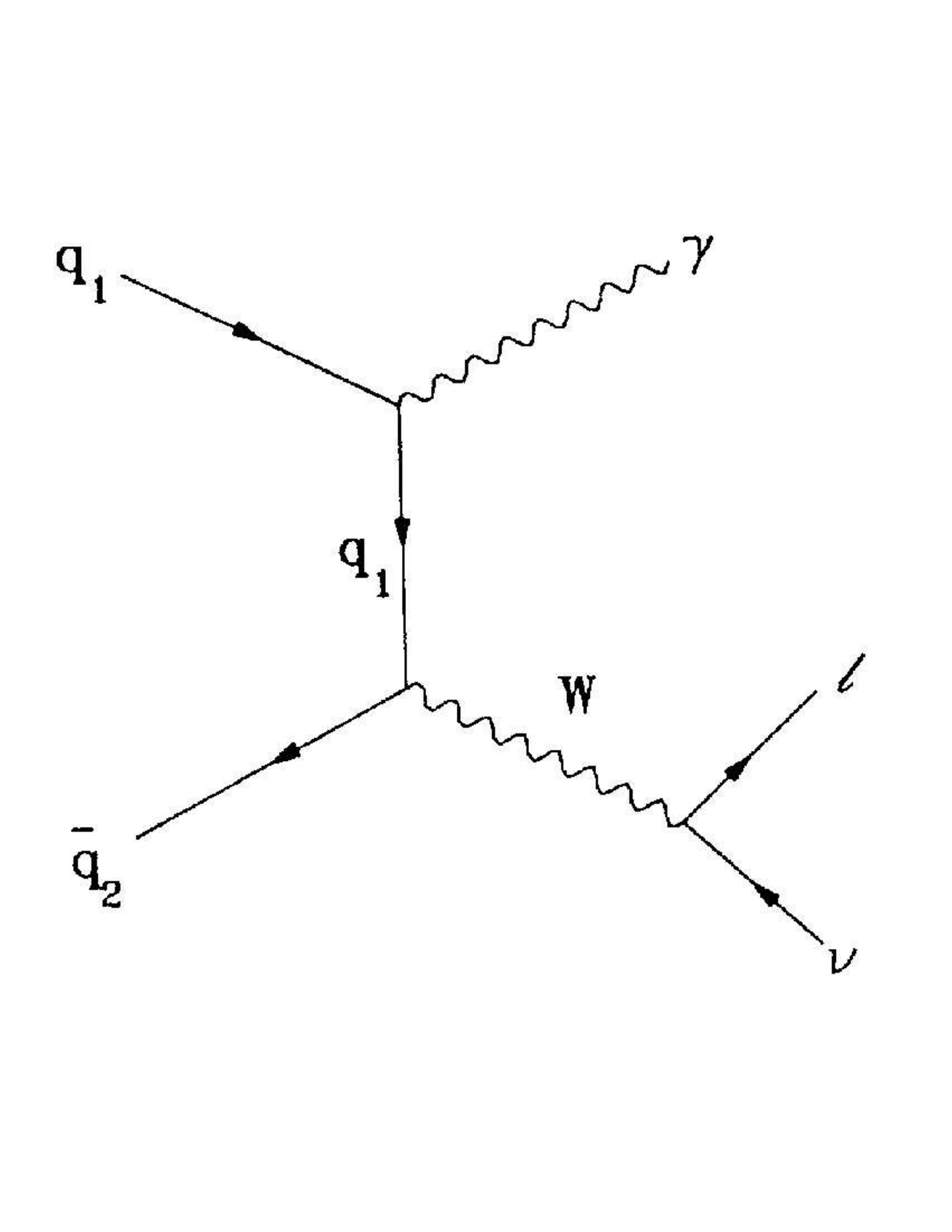} 
\caption{Born level subprocesses for \wg production in hadron-hadron collision~: s-channel process ({\bf left}), t-channel process ({\bf middle}) and u-channel process ({\bf right}).}
\label{fig:treeGraphs}
\end{center}
\end{figure}

\begin{figure}[!htbp]
\begin{center}
\includegraphics[width=0.3\textwidth,height=0.15\textwidth]{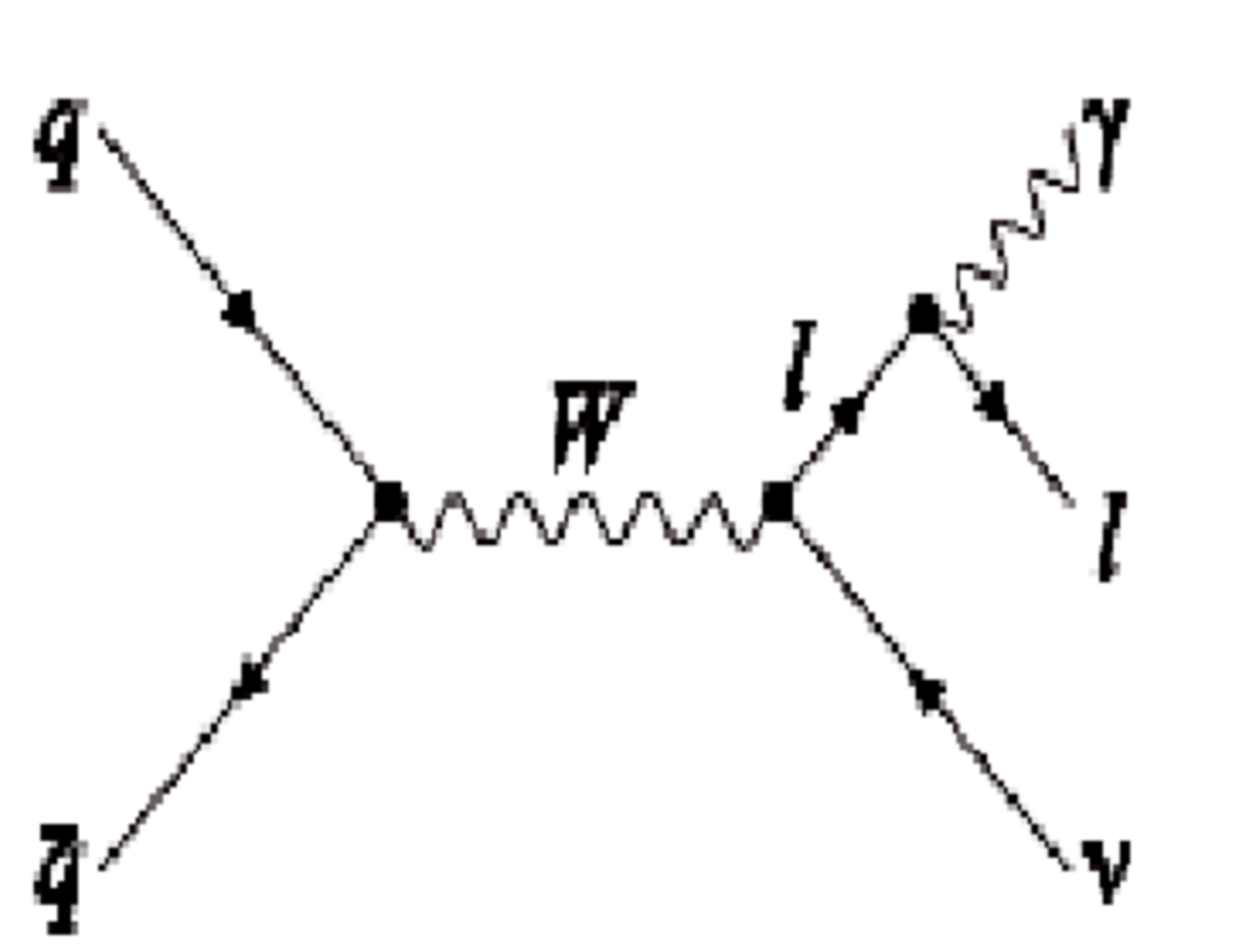} 
\hspace{0.1cm}
\includegraphics[width=0.3\textwidth,height=0.15\textwidth]{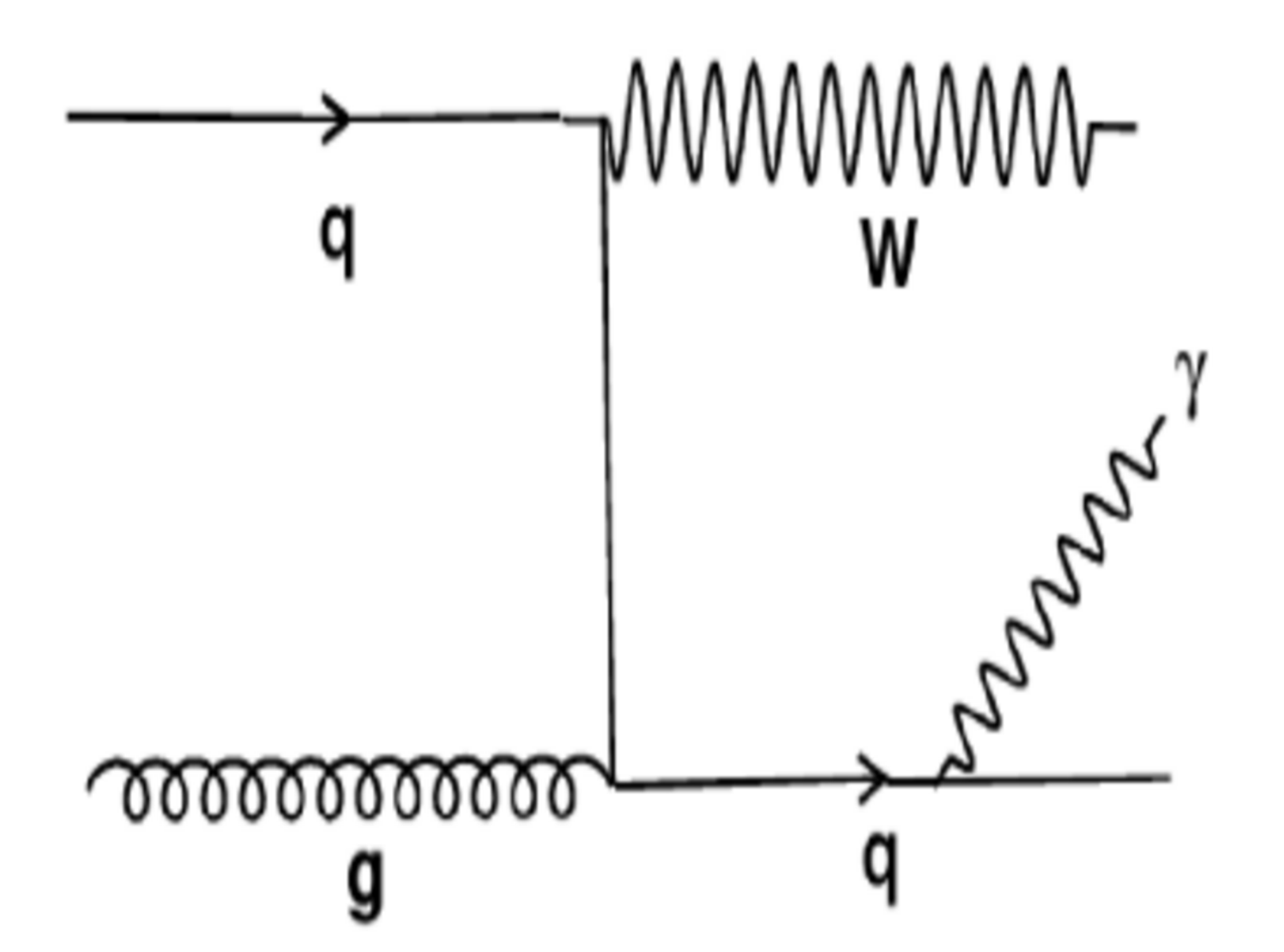} 
\caption{Final state radiation from the charged lepton from W-decay  ({\bf left}) and Quark-gluon fusion diagram producing a \wg in the final state({\bf right}).} 
\label{fig:fsr_qgFusion}
\end{center}
\end{figure}

\begin{figure}[!htbp]
\begin{center}
\includegraphics[width=0.3\textwidth,height=0.15\textwidth]{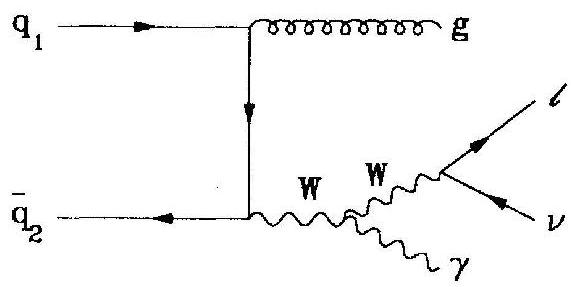} 
\hspace{0.1cm}
\includegraphics[width=0.3\textwidth,height=0.15\textwidth]{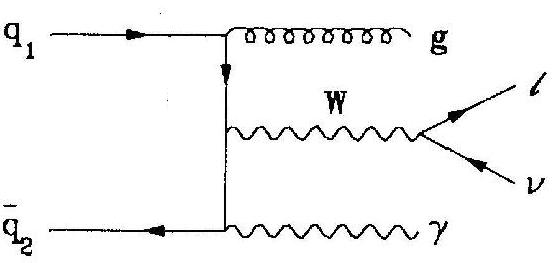} 
\\
\includegraphics[width=0.3\textwidth,height=0.15\textwidth]{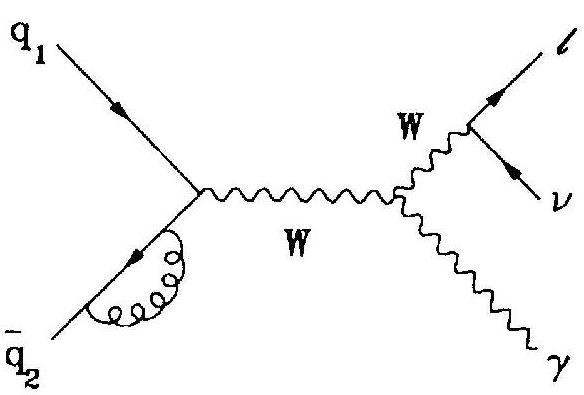}
\hspace{0.1cm}
\includegraphics[width=0.3\textwidth,height=0.15\textwidth]{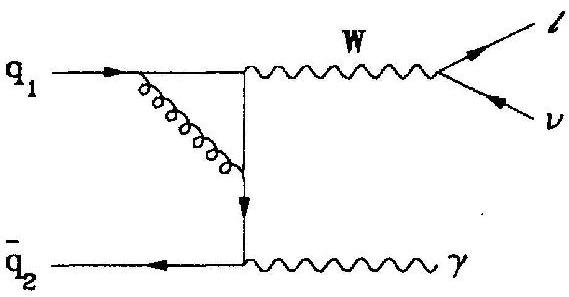} 
\caption{Higher order QCD diagrams for \wg production in hadron-hadron collision.}
\label{fig:lo_nlo_graphs}
\end{center}
\end{figure}

In hadron colliders, the Born level Feynman graphs for \wg events, with W decaying to leptons, are shown in Fig.~\ref{fig:treeGraphs}. Figure~\ref{fig:fsr_qgFusion} ({\bf left}) shown the production of W-boson which subsequently decays to a lepton and a neutrino with the lepton emitting a photon. The quark-gluon fusion process, shown in Fig.~\ref{fig:fsr_qgFusion} ({\bf right}), where an incoming quark and gluon produces a W-boson and a quark in the final state with the quark radiating a photon is also present. The FSR diagram leads to the same final state as a \wg production with W-decaying through leptons. Other leading and higher order QCD diagrams are present as well, some of which are shown in Fig.~\ref{fig:lo_nlo_graphs}, and their contribution can be quite dominant at \wg productions at high energies.
The $WW\gamma$ vertex appears explicitly in the s-channel diagram only and is discussed below. 

\subsection{$WW\gamma$ vertex and cross-section}

The most general Lorentz and electromagnetic gauge-invariant CP-conserving Lagrangian  for the $WW\gamma$ vertex can be written as follows:
\begin{equation}\label{eqn:Lwwgam}
 {\cal L}_{WW\gamma} = -ie\Bigg[W^{\dagger}_{\mu\nu}W^{\mu}A^{\nu} - W^{\dagger}_{\mu}A_{\nu}W^{\mu\nu} + \\ 
 \kappa W^{\dagger}W_{\mu}F^{\mu\nu} + \frac{\lambda}{M_{W}^{2}}W_{\lambda\mu}^{\dagger}W_{\nu}^{\mu}F^{\nu\lambda}\Bigg]
\end{equation}
where $A^{\mu}$ and $W^{\mu}$ are the photon and $W^{-}$-fields respectively, $W_{\mu\nu} = \partial_{\mu}W_{\nu} - \partial_{\nu}W_{\mu}$ and $F_{\mu\nu} = \partial_{\mu}A_{\nu} - \partial_{\nu}A_{\mu}$. The variables $\kappa$ and $\lambda$ are related to the magnetic dipole moment $\mu_{W}$ and the electric quadrupole moment $Q_{W}$ of the $W$-boson:\\
$$ \mu_{W}  = \frac{e}{2M_{W}}(1 + \kappa + \lambda), ~~~~
Q_{W} = -\frac{e}{M_{W}^{2}}(\kappa - \lambda)$$

 The vertex term in Eqn.~\ref{eqn:Lwwgam} contains both the SM and the non-Standard Model (NSM) contributions, the latter being expressible in terms of $\Delta\kappa = \kappa -1$ and $\lambda$, both of which are zero in the SM.

The differential cross-section for $q_{1}\bar{q_{2}}\rightarrow W^{-}\gamma$, in the Standard Model \cite{Brown:1979ux}, with $\kappa = 1$ and $\lambda = 0$, is given by 
\begin{equation}\label{eqn:dSigdt}
 \frac{d\sigma}{dt}(q_{1}\bar{q}_{2}\,\rightarrow\,W^{-}\gamma) = 
 \frac{\alpha}{s^{2}}\frac{M_{W}^{2}G_{F}}{\surd{2}}g_{12}^{2} 
\Bigg(Q_{1} + \frac{1}{1+t/u}\Bigg)^{2}\frac{t^{2}+u^{2}+2sM_{W}^{2}}{tu} 
\end{equation}
where $s$, $t$ and $u$ are the Mandelstam variables, $g_{12} = cos\theta_{C}$ for $q_{1}\bar{q_{2}} = d\bar{u}$ and $s\bar{c}$ and $g_{12} = sin\theta_{C}$ for $q_{1}\bar{q_{2}} = s\bar{u}$ and $d\bar{c}$. $Q_{1}e$ is the charge of the quark  $q_{1}$ and $Q_{2} = Q_{1}+1.$ The differential cross-section can also be expressed as 
\begin{equation}\label{eqn:dSigdCosTheta}
\frac{d\sigma}{dcos\theta^{*}}(d\bar{u}\,\rightarrow\,W^{-}\gamma) = \frac{1}{2}(s - M_{W}^{2})\frac{d\sigma}{dt}(d\bar{u}\,\rightarrow\,W^{-}\gamma)
\end{equation}
where $\theta^{*}$ is the angle between the $W^{-}$ and $d$-quark or equivalently between the $\gamma$ and $u$-quark in the centre of mass frame of the system.

A special feature of the gauge theory is manifested in this distribution due to the value of  $\kappa = 1$ and $\lambda = 0$ in SM. Notably, the differential cross-section $d\sigma(q_{1}\bar{q_{2}}\,\rightarrow\,W\gamma)/dcos\theta^{*}$ vanishes at a particular angle $\theta^{*}$: $cos\theta^{*} = -1/3$. This phenomenon, the {\it radiation amplitude zero} ({\it{RAZ}}), is possible only for SM values of $\kappa$ and $\lambda$. This feature can be attributed to the expression $[Q_{1} + 1/(1+t/u)]^{2}$ which vanishes for $t^{*}/u^{*} = - (1 + 1/Q_{1})$, or in other words, for $cos\theta^{*} = -(1 + 2Q_{1})$ corresponding to the charge of the quark $Q_{1}$ = $-1/3$. 

The zero occurs due to destructive interference of radiation patterns in gauge theory tree level amplitudes for the emission of massless gauge bosons \cite{Brown:1982xx},\cite{Brodsky:1982sh},\cite{Samuel:1983eg}. Any anomalous moment resulting in different values of the coupling, would destroy the occurrence of the {\it zero}. Consequently the precise measurement of the {\it radiation amplitude zero} would serve to establish the Standard Model and also to search for Beyond Standard Model (BSM) physics. The anomalous couplings can be tested in such a situation by  measuring the production rate at high value of photon transverse momentum. The D0 experiment has recently reported a study of the radiation amplitude decay at the Tevatron \cite{:2008vja}. In the present study, we deal only with the Standard Model WW$\gamma$ vertex parameters, i.e. $\Delta\kappa~=~\lambda~=~0$.

\subsection{Event generators} 

In preparation for data analysis for study of \wg events produced at the LHC, we would like to choose an event generator that has all the features of \wg production in proton-proton collision at the energy of the LHC. \wg production. The common workhorse in the CMS collaboration for generating events is the \pythia event generator. It is a general purpose generator containing matrix element calculation of many 2$\rightarrow$2 and 2$\rightarrow$1 processes as well as showering and hadronization scheme. The calculation for \wg process is present up to Born level, i.e. it includes all graphs shown in Fig.~\ref{fig:treeGraphs}. QCD radiation can be added by parton showers as well as photon bremsstrahlung from the outgoing charged lepton from W-decay using parton shower FSR. Yet, given that the probability of a hard parton emission accompanying the \wg production at the LHC and the possibility of the $WW\gamma$ vertex containing anomalous coupling parameters, we would like to choose an NLO matrix element generator with anomalous couplings. We have chosen the \baur generator and will present here a comparison of \baur with \pythia. 

\nopagebreak[4]

\section{Comparison of \textsc{Baur Wgamma\_NLO} and \textsc{PYTHIA}~6}

\subsection{\textsc{Baur Wgamma\_NLO} at the Tevatron and LHC energies} 

We first put forward a comparison of \baur at the energy of the Tevatron (1.8~TeV) and at the LHC (10~TeV). The cuts on various \baur parameters for the Tevatron is listed in Table~\ref{table:genCutTev} and those for the LHC are in Table~\ref{table:genCutLhc}. 

\begin{table}[!htbp]
\begin{center}
\caption{Generator level cuts for \baur{} for generating events at 1.8~TeV proton-antiproton collisions.} 
\begin{tabular}{cl}
\hline
Parameter & Cut \\
\hline
Photon $p_T$ & 10 GeV \\
Charged lepton $p_T$ & 20 GeV \\
Neutrino $p_T$ & 20.0 GeV \\
Jet $p_T$ & 1.0 GeV \\
Photon rapidity & 1.0 \\
Charged lepton rapidity & 2.5 \\
Jet rapidity & 10.0 \\
$\Delta R(\gamma,lepton)$ & 0.7\\
Cluster(W,$ \gamma$) transverse mass & 90 GeV \\
Soft divergence parameter & 0.01 \\
Collinear divergence parameter & 0.001 \\
Fraction of hadronic energy in a cone around the photon & 0.15 \\
\hline
\end{tabular}
\label{table:genCutTev}
\end{center}
\end{table}

\begin{table}
\begin{center}
\caption{Generator level cuts for \baur{} as used for generating LHC events at 10TeV.}
\begin{tabular}{cl}
\hline
Parameter & Cut \\
\hline
Photon $p_T$ & 5 GeV \\
Charged lepton $p_T$ & 5 GeV \\
Neutrino $p_T$ & 5 GeV \\
Jet $p_T$ & 5 GeV \\ 
Photon rapidity & 8.0 \\
Charged lepton rapidity & 8.0 \\
Jet rapidity & 8.0 \\
$\Delta R(\gamma,lepton)$ & 0.05\\
Cluster(W,$ \gamma$) transverse mass & 10 GeV \\
Soft divergence parameter & 0.01 \\
Collinear divergence parameter & 0.001 \\
Fraction of hadronic energy in a cone around the photon & 0.15 \\
\hline
\end{tabular}
\label{table:genCutLhc}
\end{center}
\end{table}

\begin{figure}
\begin{center}
\includegraphics[width=0.45\textwidth,height=0.4\textwidth]{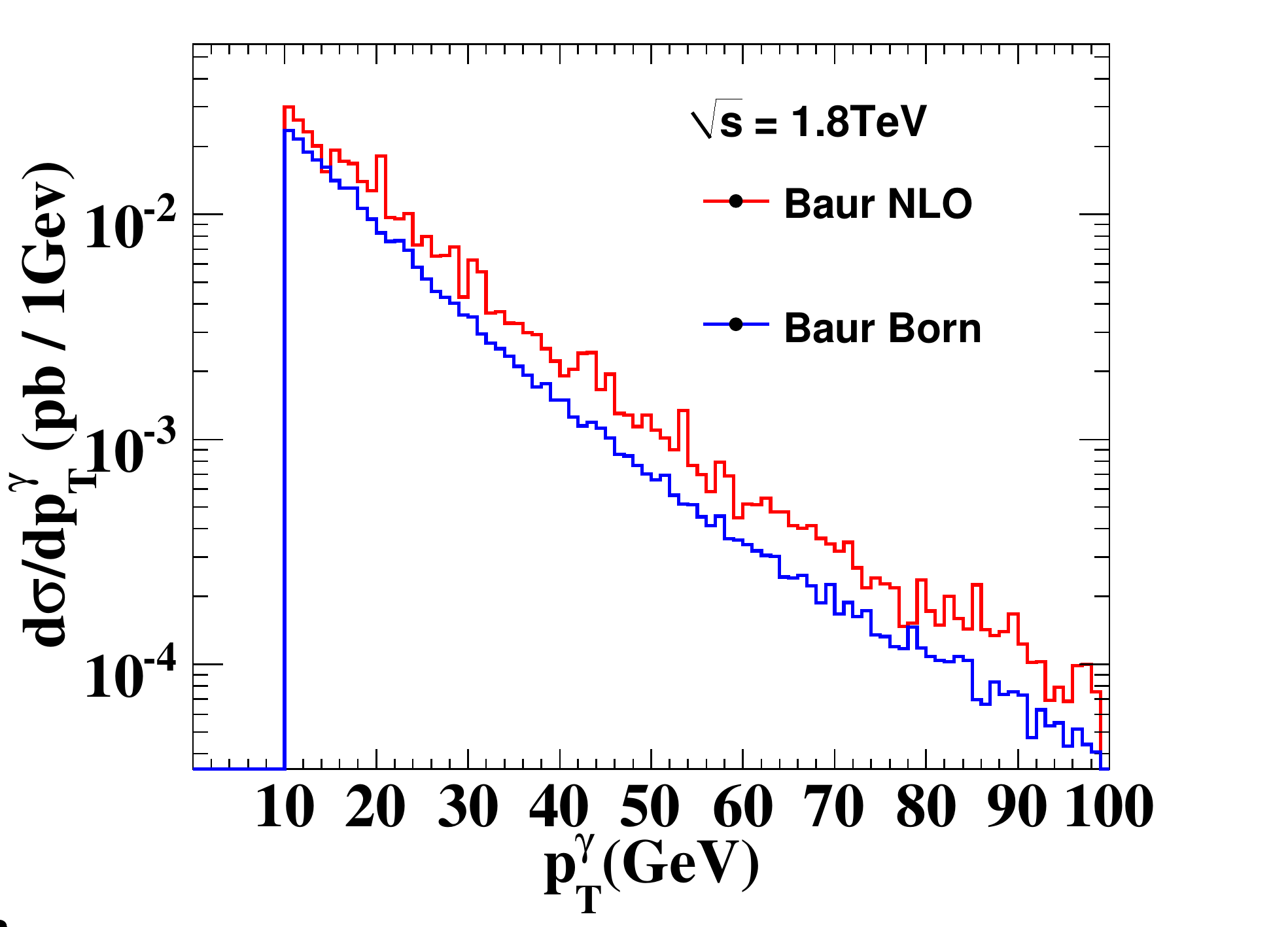}
\hspace{0.1cm}
\includegraphics[width=0.45\textwidth,height=0.4\textwidth]{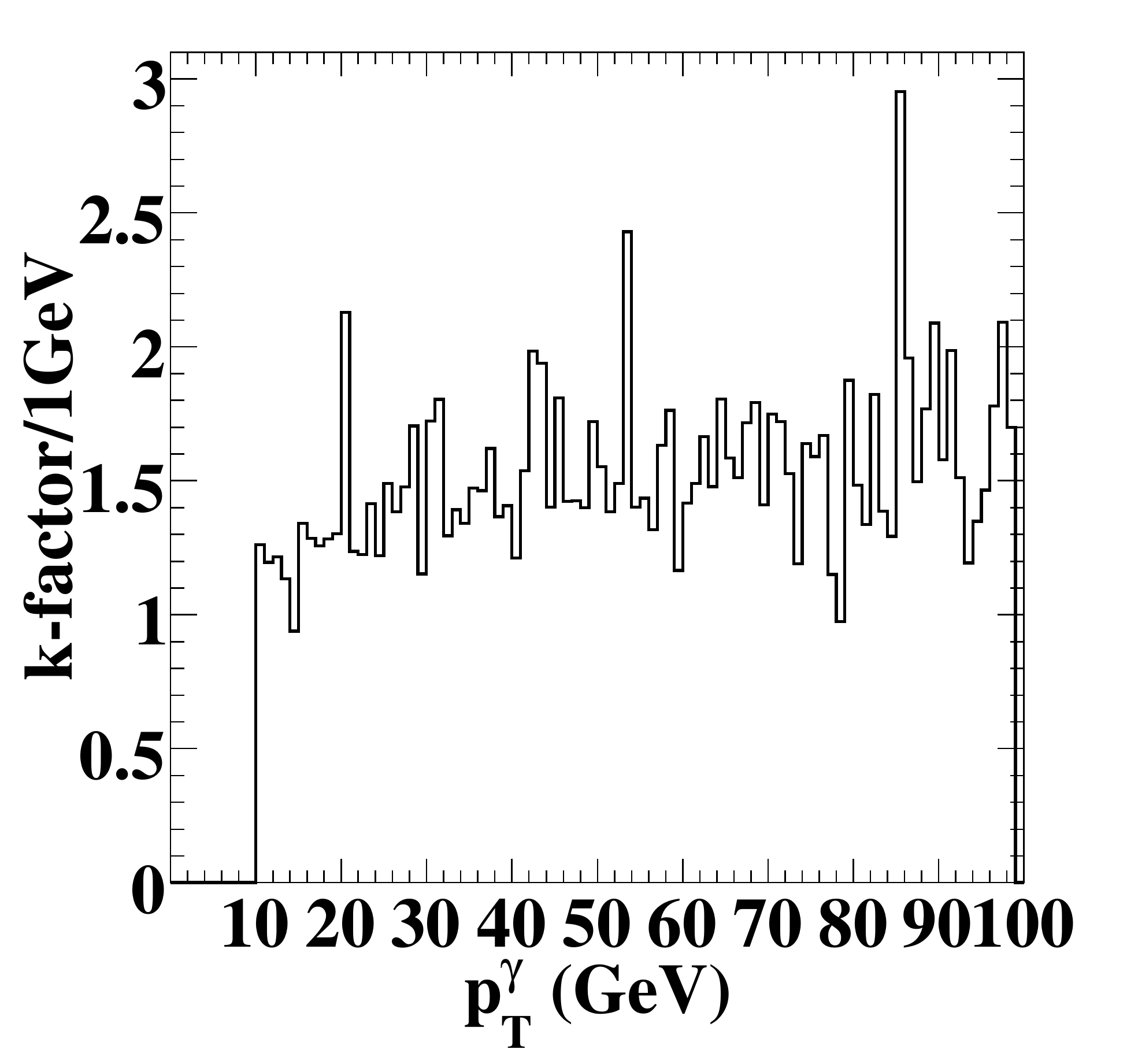} 
\caption{Photon $p_T$ from \baur at $\surd{s}$ = 1.8TeV ({\bf left}) and the k-factor, as defined in the text ({\bf right}).} 
\label{fig:bauPtgKfTev}
\end{center}
\end{figure}

Fig.~\ref{fig:bauPtgKfTev} ({\bf left}) shows the \ptg spectrum from \baur at the Tevatron. The two curves shown are for the Born level cross-section and the next-to-leading order cross-section. The k-factor is shown in Fig.~\ref{fig:bauPtgKfTev} ({\bf right}). 

\begin{figure}
\begin{center}
\includegraphics[width=0.45\textwidth,height=0.4\textwidth]{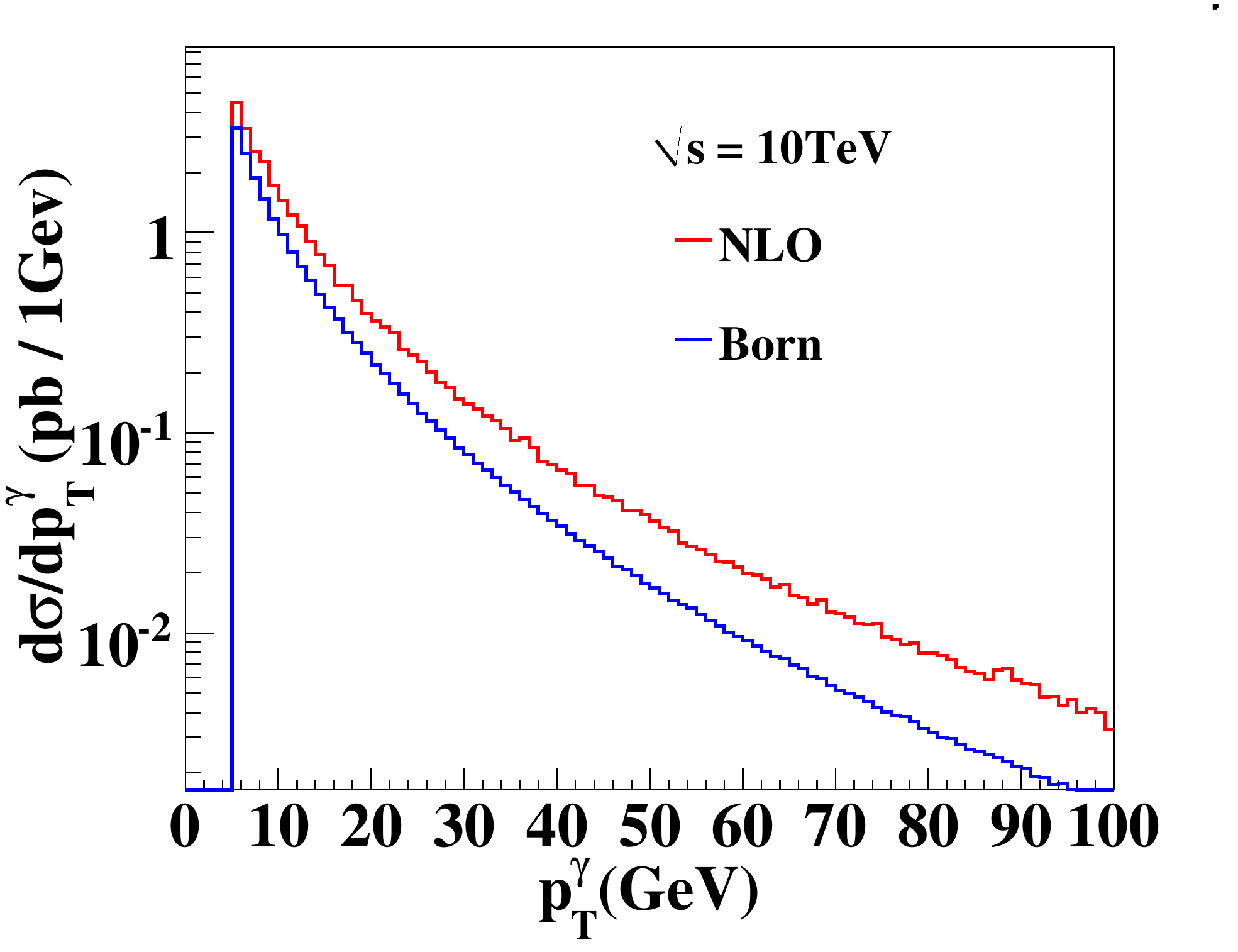}
\hspace{0.1cm} 
\includegraphics[width=0.45\textwidth,height=0.4\textwidth,angle=90.]{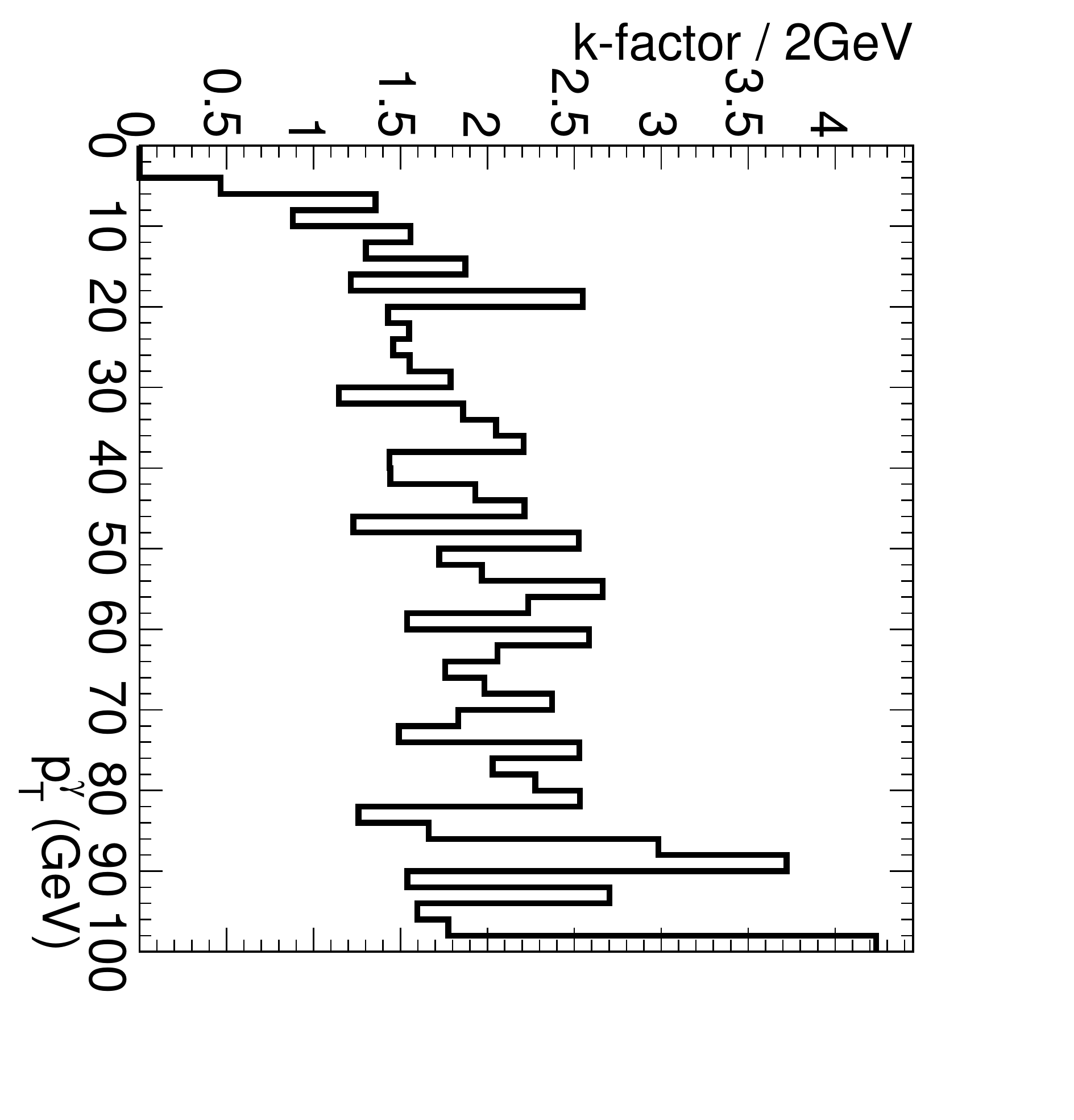}
\caption{Photon $p_T$ from \baur{} at LHC centre of mass energy, 10TeV ({\bf left}) and the k-factor at that energy ({\bf right}).} 
\label{fig:bauPtgKfLhc10}
\end{center}
\end{figure}

At the energy of the LHC, we expect an overall enhancement of the cross-section, as compared to the Tevatron energy, as shown in Fig.~\ref{fig:bauPtgKfLhc10} ({\bf left}). Further, the k-factor value (Fig.~\ref{fig:bauPtgKfLhc10} ({\bf right})) is also larger and is no longer a constant but increases with increasing value of the photon transverse momentum. Thus, for high \ptg we see that the NLO corrections become very important. 

\subsection{\textsc{PYTHIA}~6 matrix element vs \textsc{Baur Wgamma\_NLO} Born level}

We compare \baur with \pythia~6 at 10~TeV proton-proton collision. Using \pythia, we have the following options to produce \wg events~:

\begin{enumerate}
\item As mentioned before, \pythia can produce \wg events with full matrix element calculation of the Born level Feynman diagrams. We compare this \pythia 2$\rightarrow$2 process withe the events obtained from using \baur's calculation at Born level. The only difference between \baur ad \pythia at this stage is that \baur includes the decay of the W-boson to lepton in its matrix element calculation but in \pythia the W decay is treated separately. 
\item Additionally, we can generate inclusive W events in \pythia and add photon emissions using the \pythia parton showers. It is worthwhile to compare the photon transverse momentum spectrum of photons generated from the parton showers to that from \pythia's matrix element calculation and also to \baur's. 
\end{enumerate}

The \baur parameters from Table~\ref{table:genCutLhc} are used. For \pythia~6 matrix element (option 1 above), we switch on the \wg production process. ISR and FSR were switched of and so was the primordial $k_T$ of the partons inside the protons. 

\begin{figure}
\begin{center}
\includegraphics[width=0.45\textwidth,height=0.4\textwidth,angle=90.]{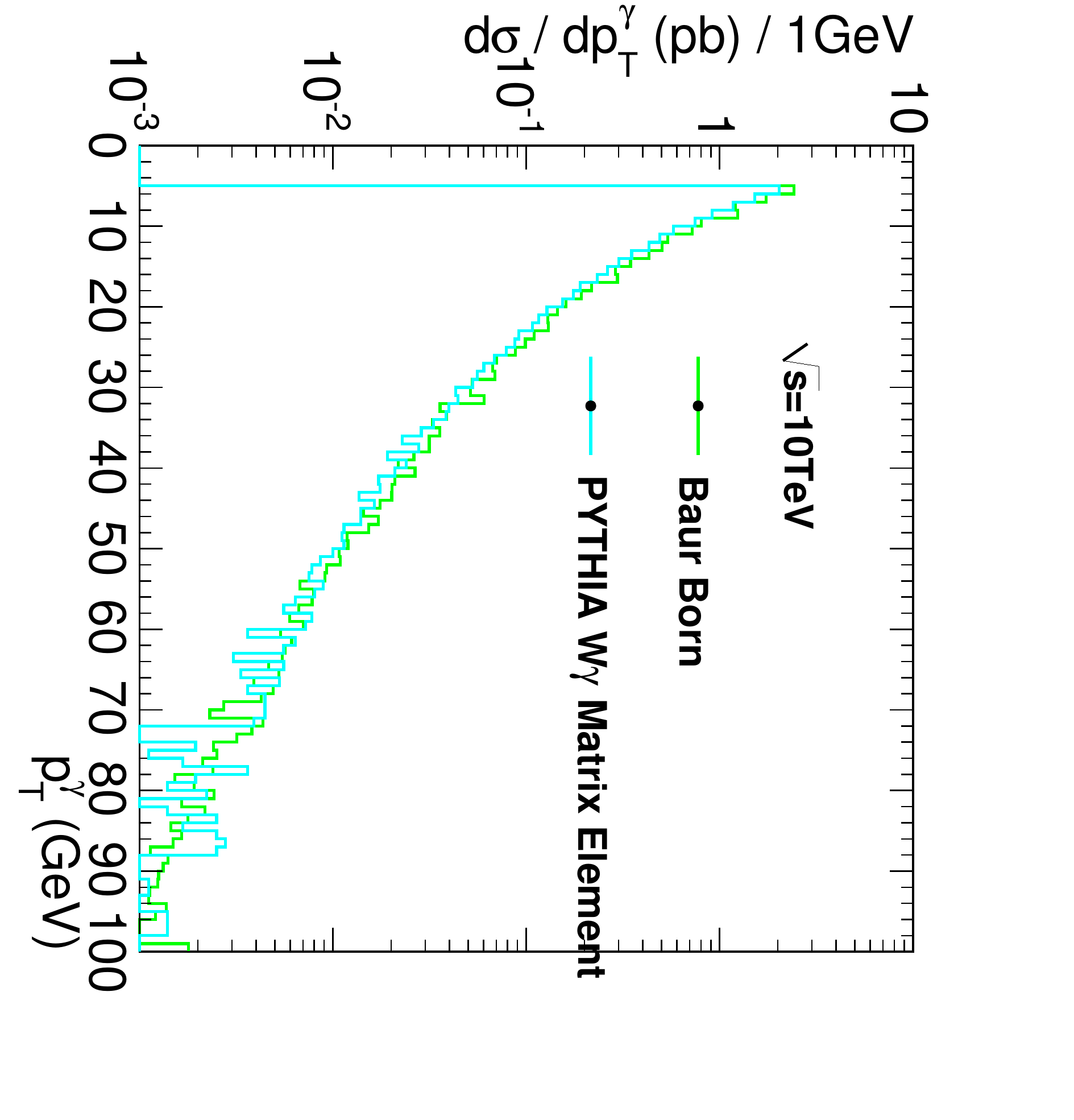}
\hspace{0.1cm} 
\includegraphics[width=0.45\textwidth,height=0.4\textwidth,angle=90.]{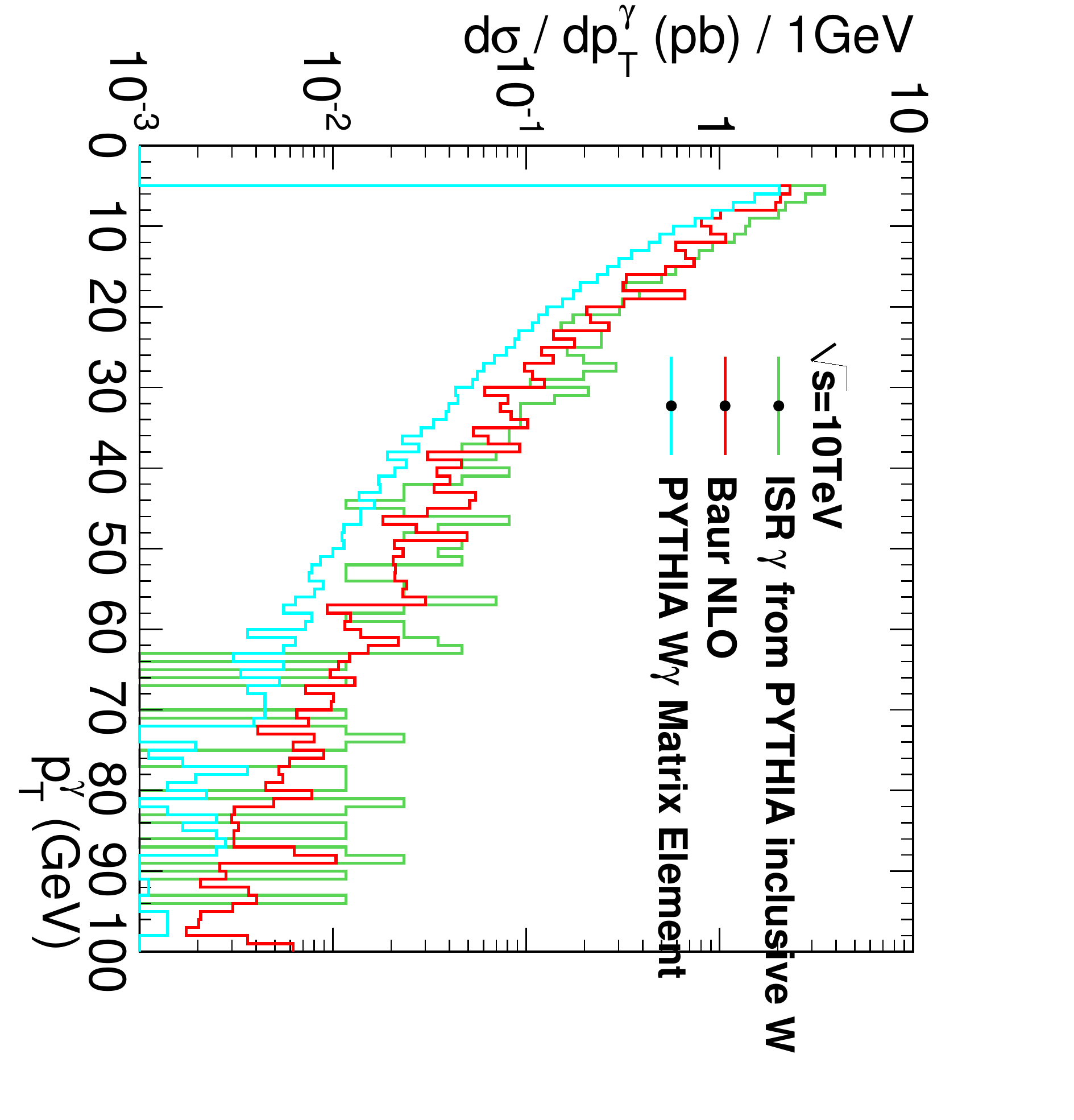}
\caption{Photon transverse momentum spectrum from \baur Born level and \pythia ME ({\bf left}) and the \ptg spectrum from \pythia \wg matrix element calculation  without PS compared to \baur NLO calculation ({\bf right}). Superimposed on the same canvas ({\bf right}) is the photon $p_t$ spectrum from \pythia ISR in an inclusive W-boson production.}
\label{fig:baur_py6_born_nlo_me_isr}
\end{center}
\end{figure}

Figure~\ref{fig:baur_py6_born_nlo_me_isr} ({\bf left}) shows the $p_{T}^{\gamma}$ spectrum from \baur Born level and \pythia matrix element. We see a good agreement between \pythia and \baur at this level. Figure~\ref{fig:baur_py6_born_nlo_me_isr} ({\bf right}) shows the $p_{T}^{\gamma}$ spectrum from \baur NLO calculation and compares it to \pythia ISR photons from parton showers in inclusive W-boson events. The $p_{T}^{\gamma}$ distribution from \pythia matrix element is also plotted on the same canvas for perspective. We see that NLO effects do play a major role in altering the shape of the photon transverse momentum distribution. However, the parton shower-generated photons mimic the spectrum from the NLO events quite well. This is because the parton showers are supposed to approximate the calculation of a process up to all orders. The only problem with the parton showers is that they are mainly designed for soft and collinear emissions and so events at the high end of the transverse momentum distribution are very infrequently generated. 

Also parton showers in \pythia~6 do not have anomalous couplings  and the radiation zero. As mentioned in Section~\ref{subsec:py8ISR}, \pythia8 parton shower also contains the radiation zero feature, included as matrix element correction to the QED component of the parton shower. 

\subsection{\textsc{PYTHIA}~6 effective "k-factor"} 

Just as the \baur generator has leading order QCD diagrams with a quark or gluon emission contribution to a boost to the \wg system, the \pythia ISR QCD radiation also serves the same purpose and contains showers up to all orders although to some approximation. This leads to defining an effective k-factor for the \pythia ISR as well which is given by the ration of \pythia cross-section for the \wg process with ISR turned on to that with ISR turned off:

\begin{center}
\begin{math}
\pythia "k-factor" = \frac{\frac{d\sigma^{ISR~ON}}{dp_{T}^{\gamma}}}{\frac{d\sigma^{ISR~OFF}}{dp_{T}^{\gamma}}} 
\end{math}
\end{center}

\begin{figure}
\begin{center}
\includegraphics[width=0.45\textwidth,height=0.4\textwidth]{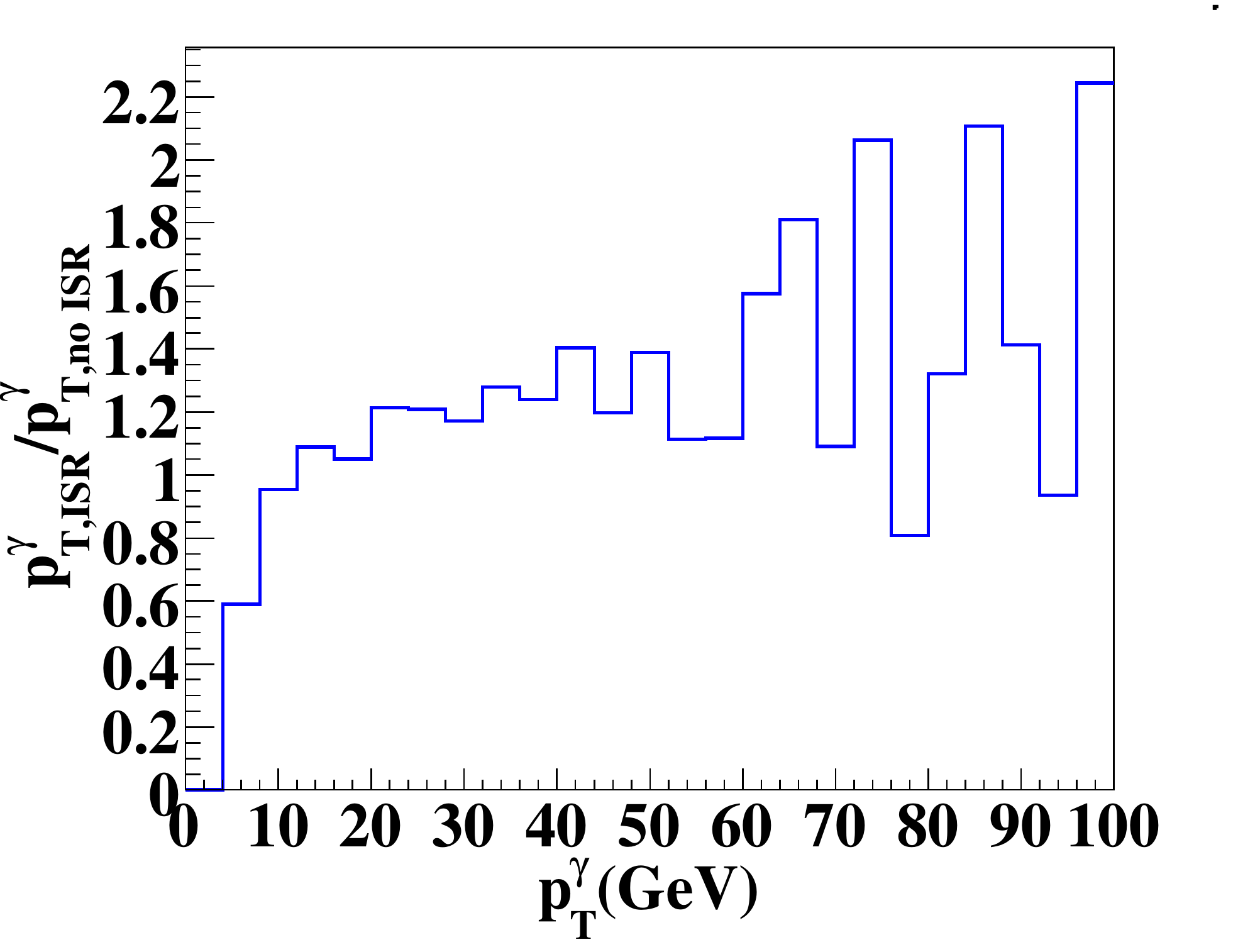}
\hspace{0.1cm} 
\includegraphics[width=0.45\textwidth,height=0.4\textwidth]{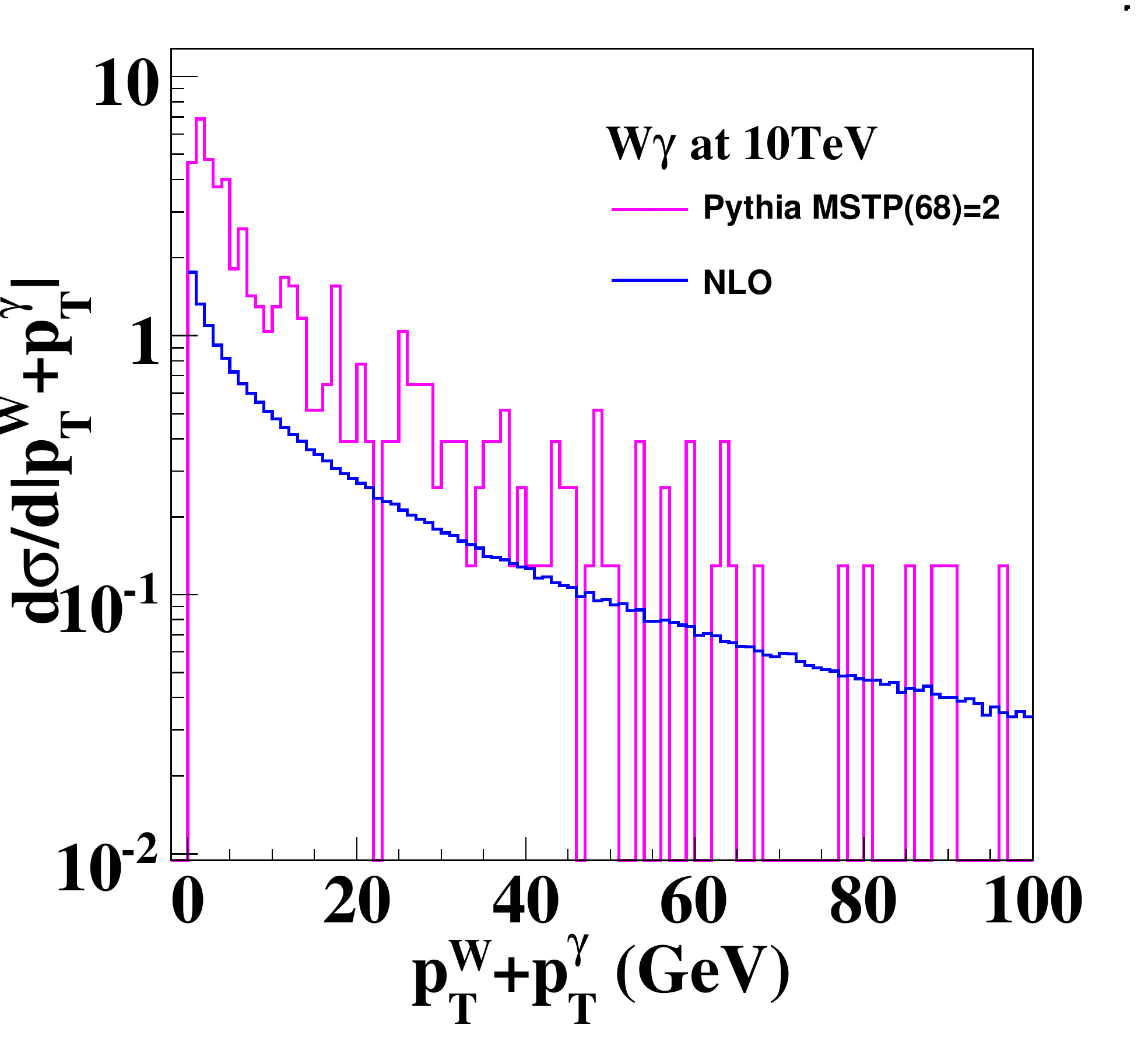} 
\caption{The ratio of \ptg spectrum from \pythia matrix element with and without ISR. This gives an effective "k-factor" for the \pythia parton shower which is supposed to take into account higher order emission diagrams for the process ({\bf left}); and the plot of the transverse momentum of the \wg system from \baur NLO and \pythia inclusive W events with ISR. This gives the amount of boost received by the W$\gamma$ from QCD radiation ({\bf right}).} 
\label{fig:pyKf} 
\end{center}
\end{figure}

Figure~\ref{fig:pyKf} ({\bf left}) shows the plot of this quantity. The ISR QCD emission, as anticipated is much softer than the matrix element calculation of \baur for a single jet emission. Figure~\ref{fig:pyKf} ({\bf right}) shows the transverse momentum of the \wg system from \baur NLO and \pythia ISR. These two plots show that the QCD radiation from \pythia~6 parton shower is comparatively softer than the hard emission from \baur matrix element calculation; the \baur k-factor, as shown in Fig.~\ref{fig:bauPtgKfLhc10} ({\bf right}) has a higher value compared to the \pythia~6 equivalent. 

\section{\textsc{PYTHIA}~6 vs. \textsc{PYTHIA}~8 and an improved parton shower description} 

With \pythia~8 \cite{Sjostrand:2007gs} , we have a a few advantages over \pythia~6. \pythia~8 gives an improved description of the parton shower description, multiple hard-scattering and a code written in C++. 

\begin{figure}[ht!]
\begin{center}
\includegraphics[width=0.45\textwidth,height=0.4\textwidth]{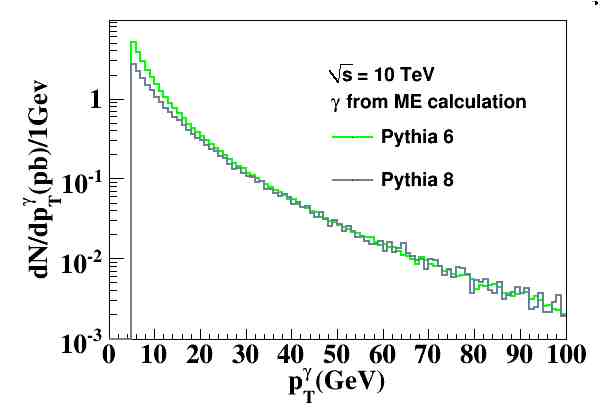}
\hspace{0.1cm} 
\includegraphics[width=0.45\textwidth,height=0.4\textwidth]{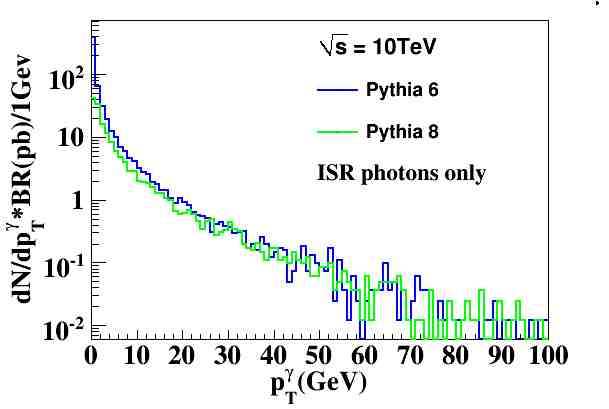}
\caption{Comparison of matrix element photon $p_T$ from \textsc{Pythia}~6 and \textsc{Pythia}~8 ({\bf left}) and comparison of parton shower photons from \textsc{Pythia}~6 and \textsc{Pythia}~8 ({\bf right}).}  
\label{fig:ptg_mePs_pyVsPy8}
\end{center}
\end{figure}

\begin{figure}
\begin{center}
\includegraphics[width=0.45\textwidth,height=0.4\textwidth]{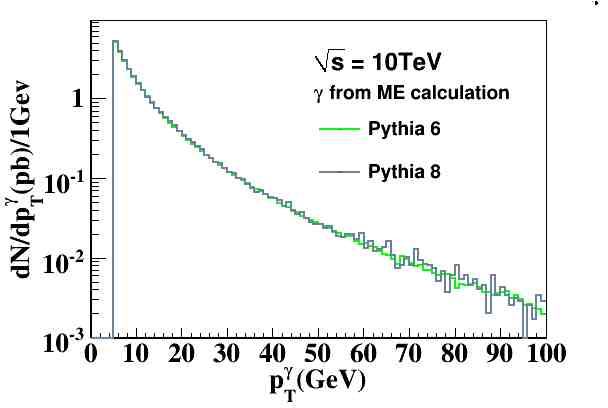}
\hspace{0.1cm} 
\includegraphics[width=0.45\textwidth,height=0.4\textwidth]{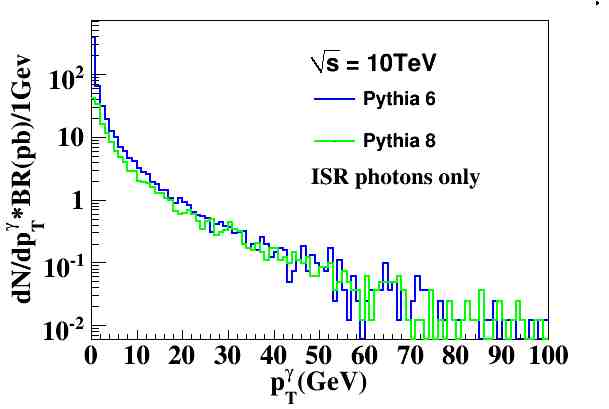}
\caption{Comparison of matrix element photon $p_T$ from \textsc{Pythia} 6 and \textsc{Pythia} 8 ({\bf left}) and comparison of parton shower photons from \textsc{Pythia} 6 and \textsc{Pythia} 8 ({\bf right}) with same settings for both \pythia~6 and for \pythia~8. We see that the agreement is better compared to that in Fig.~\ref{fig:ptg_mePs_pyVsPy8}.}
\label{fig:ptg_mePs_pyVsPy8Corr}
\end{center}
\end{figure}

The parton shower of \pythia~8 is compared to that of \pythia~6 in Fig.~\ref{fig:ptg_mePs_pyVsPy8}. Figure~\ref{fig:ptg_mePs_pyVsPy8} ({\bf right}) shows the parton shower photon $p_T$ spectrum from \pythia~6 and \pythia~8 using their default settings. The agreement is not perfect and a complete agreement between both the \pythia versions are achieved by setting the \pythia~8 parton shower setting to the default vale of \pythia~6 as shown in Fig.~\ref{fig:ptg_mePs_pyVsPy8Corr}. This comparison is important because in future we use \pythia~8 for our  matching parton showers to \baur-generated events and a consistency check is required.

\subsection{ISR, FSR and Multiple Scattering}\label{subsec:py8ISR}

\pythia~8 includes an improved description of the parton shower which agrees more closely with the NLO calculation from \baur. The parton shower photon in \pythia~8 now also contains the radiation zero which is present in the $q\bar{q}\rightarrow W\gamma$ calculation. 

Sources of photons, as mentioned before, could also be the FSR photon from the outgoing charged lepton if the W decays via leptons, as in Fig.~\ref{fig:fsr_qgFusion} ({\bf left}). At the detector level, this yields the same final state as \wg processes of Fig.~\ref{fig:treeGraphs} and so the events containing FSR need to be eliminated to select the actual events. \baur does not contain this diagram Fig.~\ref{fig:fsr_qgFusion} ({\bf left}) and \pythia adds this diagram through the parton shower evolution of the FSR. Thus using \pythia it is necessary to understand the contribution of these photons to the overall photon distributions. 

Similarly, a second hard process, e.g. $qg\rightarrow \gamma q$ may occur with a W-boson production via $q\bar{q'}\rightarrow W$ and will have all the characteristics of a \wg event. The new \pythia~8 provides the setting to study this contribution as well.

\begin{figure}
\begin{center}
\includegraphics[width=0.45\textwidth,height=0.4\textwidth]{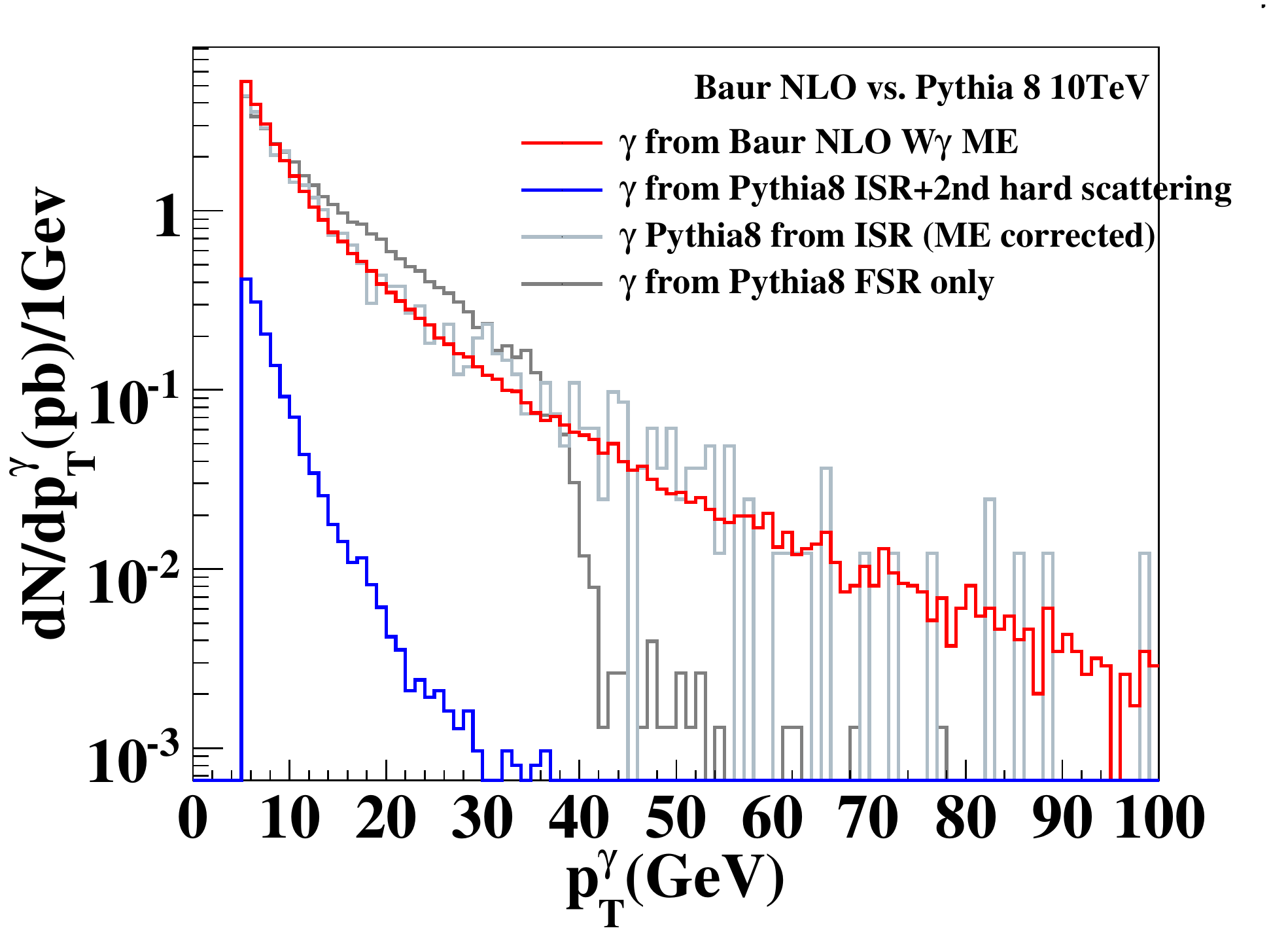}
\caption{Comparing photon $p_{T}$ from \textsc{Pythia} $W\gamma$ matrix element, from ISR shower and both ISR and a second hard interaction producing a photon in the same event.} 
\label{fig:2ndHard} 
\end{center}
\end{figure}

Figure~\ref{fig:2ndHard} shows the photon $p_T$ spectrum from both the above sources and compares it with the photons from \pythia~8 ISR and matrix element from \baur. The FSR photons do not survive above 40~GeV which is the kinematic cutoff and a reasonable cut of the \ptg on events from the detector should remove almost all of these. Interestingly, the contribution from multiple hard scattering is also insignificant compared to the photon counts from the hard scattering process and ISR and this too is confined to a low region of \ptg. 

\section{Matching}

This section is based entirely on the C++-based version \pythia~8. Consequently, any reference to \pythia would mean \pythia~8.120, unless otherwise stated. 

A collision of two protons involve a hard scattering part which produces the signal of interest as well as many other soft collisions for the other constituents of the proton, the quarks and gluons (collectively called partons), which need to be taken into account to make a complete description of the event. The debris of the collisions, both hard and soft, are never actually seen in the detector. Instead, one sees other particles which are composites of these fundamental partons which occur through he process of hadronization. This is a unique feature of the interaction that governs particles like protons and its fundamental constituents the partons. The evolution of the partons into more composite particles, known as hadrons is called hadronization and what we can detect are these hadrons. It is obvious that for claiming to model an event as a real life scenario, a Monte Carlo program needs to take care of the hadronization of quarks and gluons. Baur being a dedicated event generator for W+photons events does not take into account either the soft part of a collision nor of the hadronization. On the other hand, \textsc{Pythia} being a general purpose event generator, provides a full description of the event with the hard scattering, underlying events, parton showers adding the ISR and FSR and finally the hadronization of the products. 

We would like \baur to simulate the hard scattering two partons to produce a \wg event with W decaying to, say a muon and muon-type neutrino. After this, we will use \pythia to take care of ISR and FSR through the parton showers and then also simulate the underlying events and the hadronization of the final products. However the following problems are encountered if we naively let \pythia parton shower act upon the events from \baur: 

Since \baur is a matrix element event generator, it produces each event associated with a certain weight: i.e. it samples the phase space available uniformly and assigns a weight to each event corresponding to the probability of that region of phase space getting populated. This poses a problem because we want events to be generated with unit weights. Second, there is the problem of "double counting" in certain regions of phase space, which we discuss below:

\baur produces two kinds of events: events with no partons in the final states, with only the W and the photon with $W\rightarrow \mu\nu$ and events with a parton in the final state apart from the W decay products and the photon. The former will be called {\it 3-body events} and the latter, {\it 4-body events.} These 4-body events are the exclusive \wg+1jet events, given some suitable definition of a jet. The \baur program calculates the cross-section of both the 3-body (0-jets) events as well as 4-body (1-jet) events. 
Addition of parton showers by \pythia should preserve these cross-sections to first order. But it may so happen that during the showering of some of the 3-body events, the parton shower generated by \pythia agrees with the definition of the jets  for the 4-body events from \baur. Thus these 3-body events would be considered as 4-body (1-jet) events after the showering, thereby altering the exclusive cross-section of both \wg and \wg+1jet events. 

We have demonstrated a proof-of-principle matching scheme for Baur and \textsc{Pythia} and further details like looking that the parton spectrum and cuts on the parton (jet definition) has to be optimized. 

\subsection{The matching algorithm}

\begin{enumerate}
\item Baur WGAMMA\_NLO produces 3-body final states ($\mu$, $\nu$ and $\gamma$) and 4-body final state ($\mu$, $\nu$, $\gamma$ and jet (q or g)).
\item The 4-body events lack a Sudakov form factor (the probability of no emission which take into account the effect of virtual loops) which requires the following: 
\begin{itemize}
\item Project onto a 3-body, by assuming that the outgoing parton can be emitted from either incoming partons (flavours permitting), with relative weights by splitting kernels and parton densities. So we assume that the 4-body state never had a parton emitted and all the kinematics are recalculated based on that. This gives the projected 3-body event. 
\item Shower the resultant 3-body, compare $p_{T}^{shower}$ at the first ISR branching with the $p_{T}^{parton}$ in original 4-body.
\item If $p_{T}^{shower} > p_{T}^{parton}$ then the event is reclassified to 3-body; move to step 3 below.
\item If $p_{T}^{shower} < p_{T}^{parton}$ then the original 4-body event is  now showered and $p_{T}^{shower}$ is compared with $p_{T}^{parton}$.
\item If $p_{T}^{shower} > p_{T}^{parton}$ then go back one step.
\item Continue with rest of shower to give a complete event.
\end{itemize}
\item Shower the 3-body events; compare $p_{T}^{shower}$ with $p_{T}^{separate}$ after first ISR branching. $p_{T}^{separate}$ can be considered to be the boundary between the ME calculation's regime and that of the parton shower's. 
\item If $p_{T}^{shower} > p_{T}^{separate}$ then stop any further shower evolution and go back one step.
\item Continue with the rest of the shower.
\end{enumerate}

\subsection{Matching results}

According to the above-mentioned algorithm, we first project 4-body states on to 3-body states and try to determine the Sudakov form factor for the QCD emission which is originally from the matrix element calculation. 

\begin{figure}
\begin{center}
\includegraphics[width=0.45\textwidth,height=0.4\textwidth]{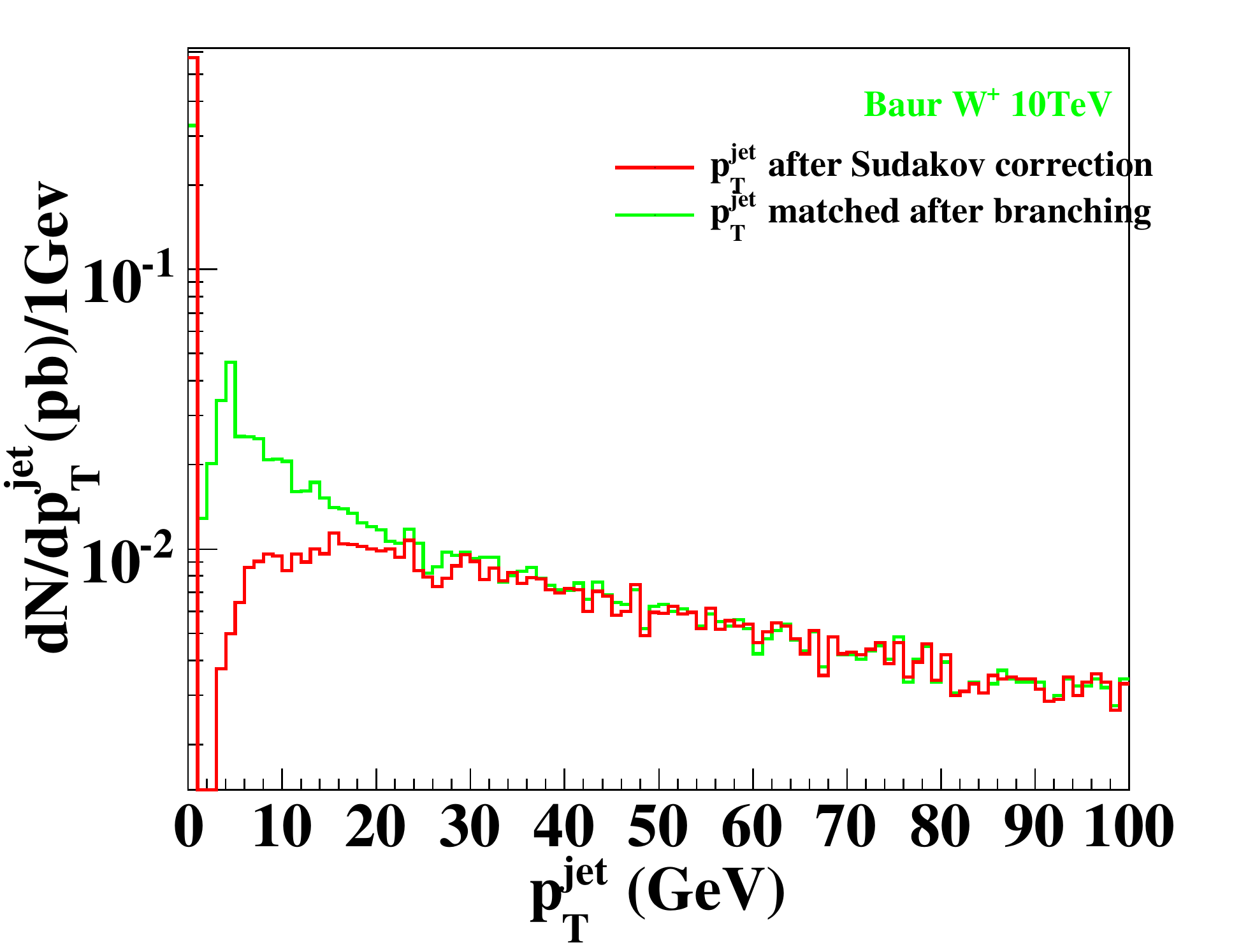}
\hspace{0.1cm} 
\includegraphics[width=0.45\textwidth,height=0.4\textwidth]{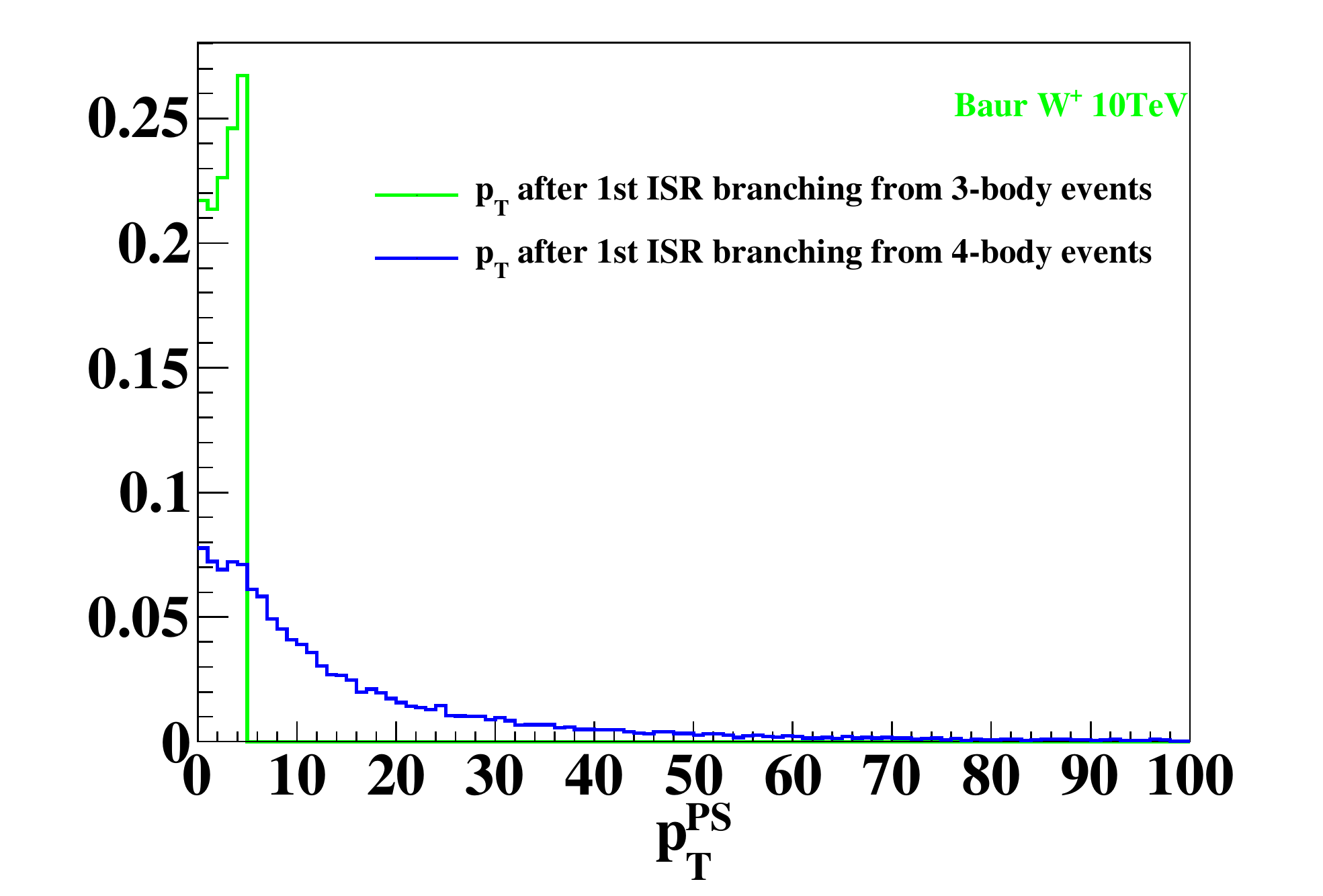}
\caption{Transverse momentum of the partons after Sudakov correction and matching ({\bf left}). Events which were originally of the type 4-body and subsequently gets reclassified as 3-body after the Sudakov correction end up with no partons in the final state. These events show up as the delta-function peak in the 0th bin of the red histogram. The Transverse momentum of the partons from 3 and 4-body events after 1st ISR emission are shown on the {\bf right}.} 
\label{fig:3-4body}
\end{center}
\end{figure}

Figure~\ref{fig:3-4body} ({\bf left}) shows the transverse momentum of the jets after the Sudakov correction and the $p_T$  of the jets matched after branching. The delta-function at the origin correspond to those events which are reclassified as 3-body events after the Sudakov correction. 

As a second step to the parton matching scheme, the events are showered and the first ISR $p_T$ is plotted in fig~\ref{fig:3-4body} ({\bf right}). Since according to our algorithm, we are clearly demarcating a $p_T$ region above which matrix element calculation is valid and below which is the parton shower regime (at 5~GeV for the present study) we see that the first ISR emission from 3-body events are always confined to below 5~GeV. But this is not so in the case where the parton showering is only required to be softer than the matrix element parton. Hence for 4-body events, we get a long tail for events with highly energetic matrix element parton.

\begin{figure}
\begin{center}
\includegraphics[width=0.45\textwidth,height=0.4\textwidth]{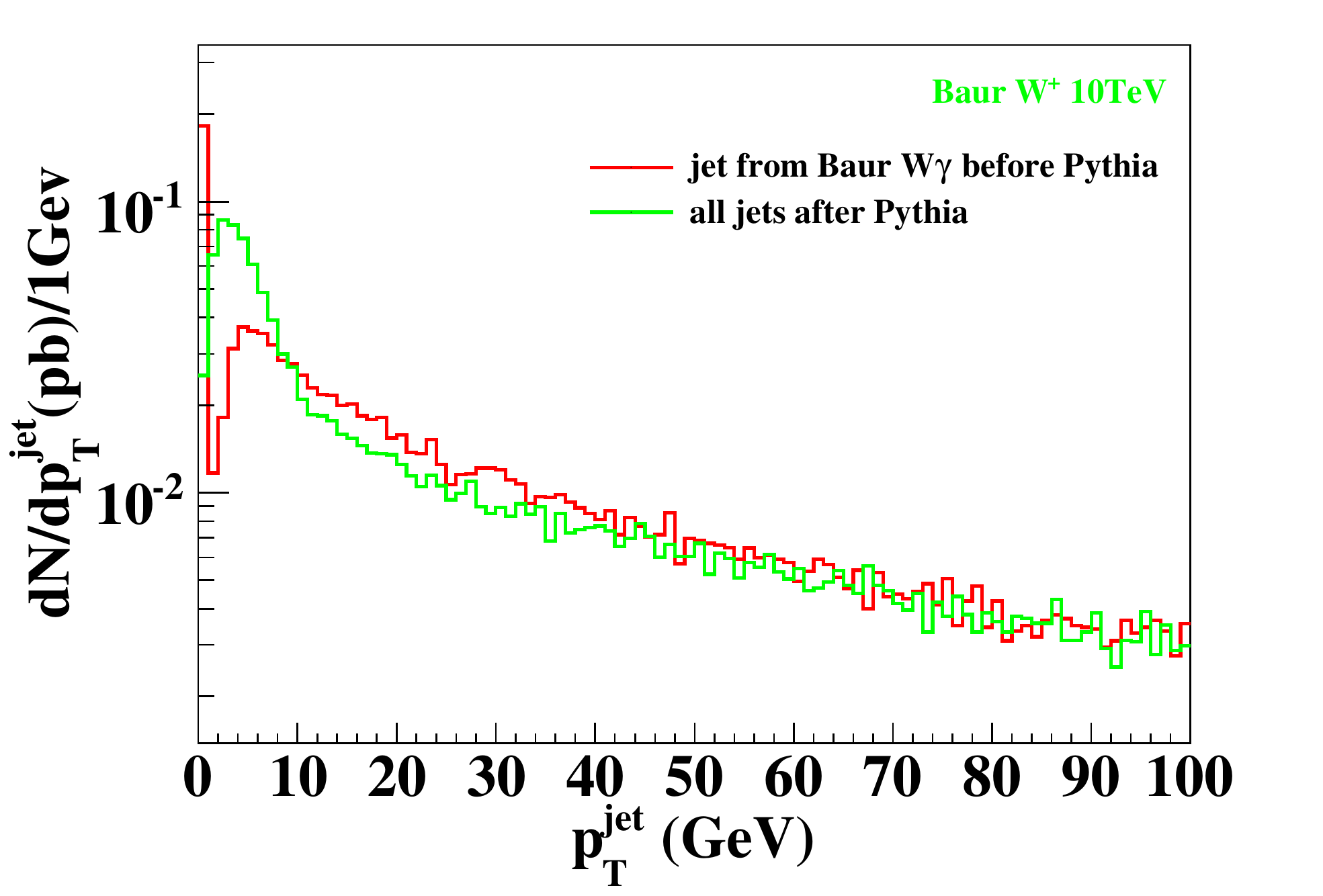}
\label{fig:alljets}
\caption{Transverse momentum of the \wg system before \pythia showering and after \pythia showering.} 
\label{fig:wgPt}
\end{center}
\end{figure}

Figure~\ref{fig:wgPt} shows the shows the distribution of the transverse momenta of the \wg system as produced by \baur and after the \pythia showering. The red histogram, depicting the distribution from \baur shows many events with $p_{T}^{W\gamma}$ equal to zero which are the 3-body events. The non-zero values correspond to events with a parton in the final state. In the green histogram, we see that the kink after the zeroth bin fills up due to the boost from \pythia ISR. The area under both these curves however remain the same indicating that the exclusive cross-section of the 1-jet events remain conserved after the parton shower. 

While being suitable for a simple event topology with one jet, the matching scheme does not require any modification of the matrix element calculation of Baur for the Sudakov form factor. 

\section{Summary }

The preparation for data analysis of high energy physics experiments involve the use of Monte Carla techniques to simulate generation of events one wishes to study and also simulation of detector equipment and interactions of the produced particles in the detector. As a preparation for studying the production of W$\gamma$ events at LHC proton-on-proton collision using the CMS detector, we wish to accurately model the production mechanism of W+$\gamma$ at the LHC. In this note, we have made a comparison of a well-established general purpose event generator, \textsc{Pythia} with the dedicated \baur generator. \textsc{Pythia} has been found wanting on several issues regarding \wg production mechanism and \baur too has been found to lack the total description of a realistic hadron collision scenario leading to \wg production. The most effective strategy, thus would be to combine the two event generators to the best of our advantage and in the later section of this note, we have outlined a home-brewed approach to this combination, the so called ``Matching strategy''. Given the individual performance of Baur and \textsc{Pythia}, a successful implementation of the matching strategy will pave the way for event generation taking into account aspects of all physics involved, at various scales, in hadronic production of \wg at the LHC. 

\bibliography{baur_pythia}

\end{document}